# Estimating Commuting Patterns from High Resolution Phone GPS Data

Bita Sadeghinasr[1], Armin Akhavan[1], Qi Wang, Ph.D.[1]

[1] Department of Civil and Environmental Engineering, Northeastern University, 360 Huntington Ave, Boston, MA 02115
*Correspondence: q.wang@northeastern.edu

**ABSTRACT**

The rise of location positioning technologies has generated enormous volumes of digital footprints. Translating this big data into understandable trip patterns plays a crucial role in estimating infrastructure demands. Previous studies were unable to correctly represent commuting patterns on smaller urban scales due to insufficient spatial accuracy. In this study, we investigated if, and to what extent, estimated commuting patterns identified from GPS data can replicate the results from transportation surveys and to what degree these estimates improve the estimates of trips distribution pattern on census tract level using higher resolution data. We inferred average daily home-to-work trips by analyzing phone GPS data and compared these patterns with U.S. Census summary tables. We found that trips detected by GPS data highly correlate with census trips. Furthermore, GPS data is a better proxy for Census tract-pairs with larger numbers of trips.

**INTRODUCTION**

Gaining a profound insight into human mobility is of crucial importance to many areas such as urban planning (Appleyard et al. 1964, Hägerstraand 1970, Carlstein et al. 1978, Jiang et al. 2012), emergency response and evacuation (Wang et al. 2014, Wang et al. 2016), traffic monitoring and travel demand forecasting (Wilson et al. 2004, Treiber et al. 2013). Human movements and activities in conjunction with infrastructures' sprawl shape urban patterns; therefore, comprehending and predicting urban movement patterns can enormously contribute to finding solutions to urban complexities. The importance of broadening understanding of urban mobility has required planners to seek different sources of information on this subject (Fan et al. 2008). Traditionally, planners and policymakers benefit from household travel surveys and census data to learn about people's whereabouts. However, these surveys are time-consuming and expensive to conduct (Meyer et al. 1984, Stopher et al. 2007). Additionally, surveys only capture a snapshot of the travel behavior of a sample of people and are susceptible to self-reporting and upscaling errors (Palmer et al. 2013).

      A fruitful direction of urban mobility study is to estimate trip patterns. Having an accurate estimate of daily trips can facilitate traffic congestion management and travel time forecasting. Trips can be constructed by estimating origins and destinations (ODs). Calculating OD matrices enables authorities to better estimate volumes of traffic in transportation networks and develop the infrastructures accordingly (Barbosa et al. 2018). Among trips with different purposes, commute trips account for the largest portion of travels during peak hours (Polzin et al. 2015). Therefore, having a clear picture of the distribution of these trips plays a key part in managing traffic demand. ODs are traditionally estimated using travel surveys. Yet, due to travel surveys' shortcomings,

there have been efforts to estimate trips seeking other datasets and practices (Pan et al. 2006, Caceres et al. 2007, Sohn et al. 2008).

Previous studies have most prevalently adapted call detail records (CDR) for inferring trip patterns due to the availability of CDR on millions of users (Barbosa et al. 2018). Many studies have estimated commute trips using CDR data and validated them against travel surveys (Zhang et al. 2010, Calabrese et al. 2011, Frias-Martinez et al. 2012, Alexander et al. 2015, Toole et al. 2015, Jiang et al. 2016). They assigned OD matrices generated from CDR to networks of roads and estimated average daily OD trips by purposes and time of day. They compared estimated CDR trips with travel survey datasets and found that there is a strong correlation on the town level.

Although CDR has provided researchers with useful insights into trip patterns (Blondel et al. 2015), using CDR to detect trips between smaller geographic levels such as tracts can be problematic, due to the fact that users' locations are approximated by the position of the tower that their cell phone is connected to. Coverage area by cell towers considerably varies from tens of meters in the densest areas up to a few kilometers in rural areas (Calabrese et al. 2011). In less urbanized areas, users might transmit all communication through one tower while moving in the area covered by the same tower. However, in dense areas, users ping multiple towers with much smaller movements (Barbosa et al. 2018).

GPS data provides researchers with a high level of spatial accuracy and temporal frequency and thus can be a rich source for detecting mobility patterns (Barbosa et al. 2018). Previously GPS datasets were most commonly collected from smaller groups of individuals (Calabrese et al. 2011, Blondel et al. 2015) or trackers on cars (Bazzani et al. 2010, Pappalardo et al. 2013) and often were not available at the scale of a city. Thus, commuting patterns had seldom been explored using complete GPS datasets. Recently, data sets generated by phone GPS are emerging and have been used for mobility research (Li et al. 2008, Zheng et al. 2008, Zheng et al. 2009, Zheng et al. 2010, Akhavan et al. 2018). As opposed to travel surveys that report respondents' travel behavior over the past single day or a few days, using technologies like GPS enables us to capture travelers' behavior for many days possible. However, before using them to estimate urban trips, especially commuting trips, the limitations of the dataset should be quantified. The objective of this study is to investigate if, and to what extent, the estimated commuting patterns identified from phone GPS data of millions of users can replicate the results from commuting surveys. In order to do so, we aim to estimate daily trip patterns using ODs extracted from millions of phone GPS records.

## DATA

In this study, we analyzed 810 million phone GPS records from 1 million users in Houston, Texas over a course of two consecutive weeks (1st-15th August 2017). Phone GPS data was generated by more than 50 mobile applications that anonymously collect users' geolocation records. Each record has an anonymous device ID, latitude, longitude and time stamp.

## STUDY AREA

The area of study is the city of Houston. The Census Bureau follows people for determining census tracts rather than political boundaries. As a result, census tracts don't necessarily follow city boundaries (1994). Most of the census tracts are located entirely within the city boundaries, but some cross over the city limits (Figure 1). In order to accurately capture the trips of Houston residents, we set the limit of our analysis to tracts in which 50% or more of the area was contained in the city of Houston limits.

**METHODOLOGY**

**Stay extraction.** In order to infer trips and activities from GPS data, we need to detect areas where the user has remained stationary for a while. In this study, these areas are referred to as "stay points". Finding stay points helps us detect significant places where the user has engaged in an activity and filter out GPS noise caused by user dependent and independent errors.

In this study, we developed a method based on the work from (Li et al. 2008) to detect stay points. In order to eliminate temporary recurring stay points, such as routine stops at road intersections, time and distance were both used as filters. Each stay point is marked by its latitude, longitude, arrival time and departure time. Stay points' latitudes and longitudes are computed based on the mean coordinates of spatially-temporally close group of GPS points. The arrival time and departing time are associated with timestamps of the first and last temporally ordered points in the group of points respectively.

We detected stay points using this method rather than clustering GPS points due to the followings. First, conventional clustering methods such as Density-Based Spatial Clustering of Applications with Noise (DBSCAN) and Hidden Markov Models cause chaining effects, which reflect routes of travel and created spread out clusters; this would distract from the goal of finding centralized stay points. Furthermore, clustering raw GPS records merely based on the geographies would dismiss their temporal sequence. On the other hand, to distinguish short visits from more significant places like homes or workplaces, we introduced time thresholds in forming stay points to account for the time that user spent in those areas. In our experiment, an area was selected as a stay point if the user exceeded the time threshold within a radius of 250 meters.

**Stay region.** After extracting stay points, hierarchical clustering with complete linkage was performed to cluster stay points that are spatially close. All points belonging to one cluster were within 250 meters from each other. The center of these stay point clusters is referred to as "stay region". The number of visits to a stay region is counted by the number of stay points within each cluster. This is due to the fact that stay points are spatially close to each other but may have occurred on different days. For a given number of visits, longer trips are more likely to be the workplaces.

**Home location.** In order to produce home-to-work (HW) trips, we need to detect individuals' home and workplaces. In this study, home locations were assumed to be places where users typically spend time during night hours (8 pm- 5 am) with the greatest number of visits. Therefore, to detect homes, stay points were chosen with (1.) night hours of at least 3 hours or (2.) stay duration of more than 24 hours. These were hierarchically clustered into stay regions. Both weekends and weekdays were used to detect home places.

Table 1. Summary of GPS user statistics

|  | Individuals | Percentage |
| --- | --- | --- |
| Users in the raw data set | 1,000,000 | 100% |
| Users with "home" detected | 286,718 | 29% |
| Users both "home" and "work" detected (commuters) | 40,623 | 14.1% |

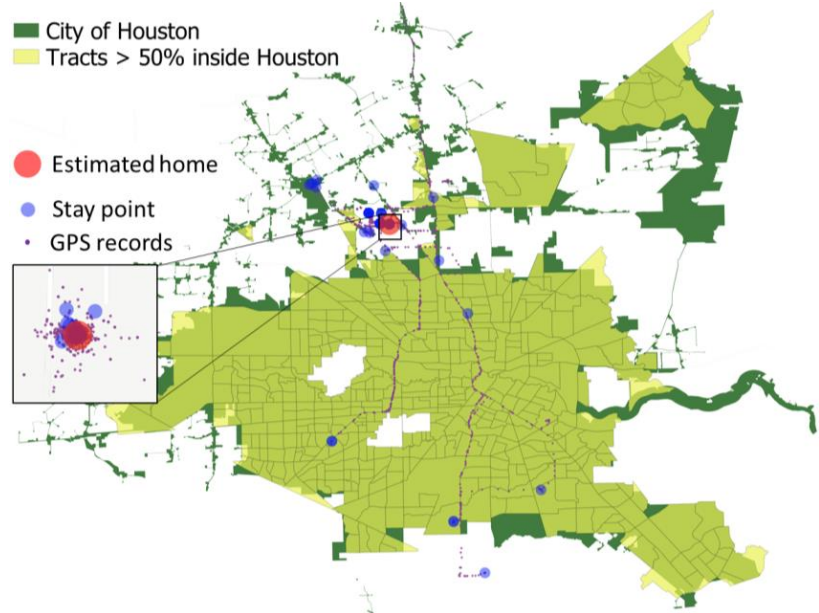

Figure 1. The city of Houston boundaries and tracts with more than 50% area inside city boundaries. Stay points and estimated home location of one individual

**Work location.** To detect work locations, stay points with arrival time within working hours (8am-6pm) on weekdays were clustered. Based on previous studies (Levinson et al. 1994, Schafer 2000) for a given number of visits, longer trips are more likely to be the work places. Therefore, for potential workplace clusters attributes of "distance from home" ($d$) and "number of visits" ($n$) were computed. Clusters within walking distance (half a mile ~ 800meters) from home and with visits less than two times during the studied period (on average once a week) were dismissed. Finally, clusters with the biggest values for ($n \times d$) were selected as work locations. We also examined $n^2$ and $n^3$ in order to counter skewed clustering as the results of the long-distance travels. However, we found that less than two percent of the users have different "work locations" identified by higher powers of $n$.

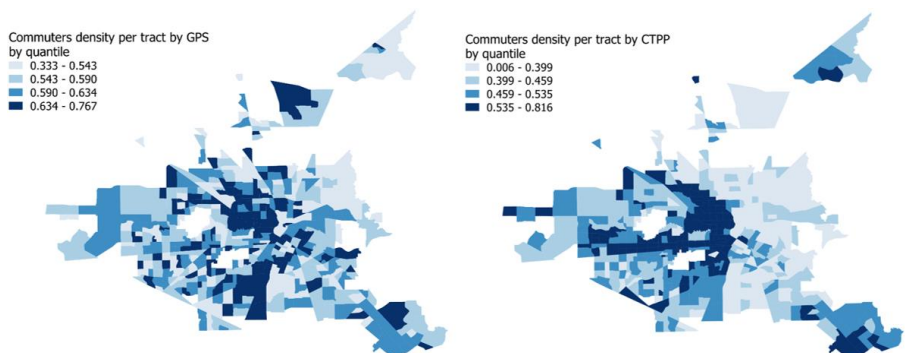

Figure 2. (a) Commuters' home distribution in tracts of study area by GPS. (b) Commuters' home distribution in tracts of study area by CTPP data.

**Average daily home-to-work trips.** To account for the fact that not all commuters go to work on every weekday, average daily HW trip was computed for each commuter. It was presumed that on weekdays users start their trips from home. Therefore, we assume that a commuter went to work if they had a stay point in a walking distance (800(m)) from their work location. In such case, their trip origin was assumed to be their home location even though there was no record close to her home location. The number of days that they commuted were then divided by the number of weekdays for each individual. Lastly, these average daily trips were aggregated into pairs of census tracts.

**Commute distance and duration.** An open source routing engine was used to estimate the commute duration and distance for each individual. The data was requested as historical data for morning peak through Navigation and Routing API of Here Technologies, for both personal cars and public transportation. The API is fed with a pair of origin and destination [home, work] and returns corresponding estimated travel distances and durations.

**RESULT AND DISCUSSION**

To validate the distribution of the commute flows detected by GPS phone data, we compared them to flows between an individual's home and workplaces reported by 2006–2010 Census Transportation Planning Products (CTPP). The correlation between the estimated commuting trips from the GPS phone data and the CTPP tract-pair home-work trips was 0.61 with a high level of significance (*p-value* < 0.0001), which shows that GPS phone data can be used to estimate urban commuting patterns. Additionally, this correlation indicates that our sample of mobile phone users can accurately explain the distribution of trips between census tracts. Figure 3 shows that relationship seems more linear for tracts where the census estimates are larger for tract-pair trips and GPS trips validate better with them. This trend can be explained due to the sparse nature of the data in tracts with a small number of commuters. Furthermore, survey data is susceptible to sampling and upscaling errors, and the uncertainty of the number of trips between tract pairs with a few numbers of commuters is reflected in the relatively large values of standard errors that are reported along with estimated home-to-workflows by CTPP. Yet, using two weeks of high-resolution GPS data holds higher correlation with CTPP flow by a factor of 17% at the tract level, as compared to the previous study (Alexander et al. 2015) using CDR data over a period of two months.

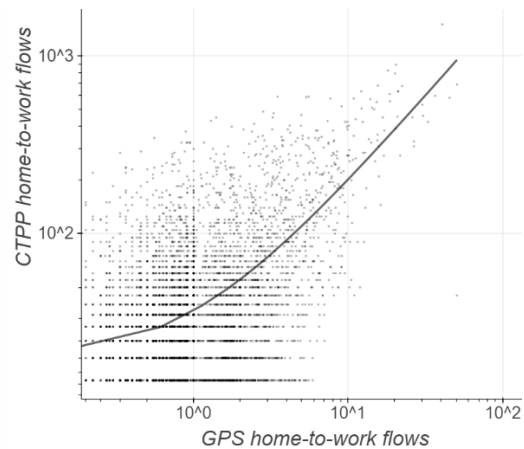

Figure 3. Correlation of tract-pair trips reported by CTPP and GPS data.

According to the routing engine, the estimated mean commute time by GPS data using personal cars was 27 minutes, which is close to 26.8 minutes estimated by the census, with 88% of the sample commuting in driving mode. Mean commute distance was estimated to be 22 km for people driving personal cars and 15 km for commuters taking public transits.

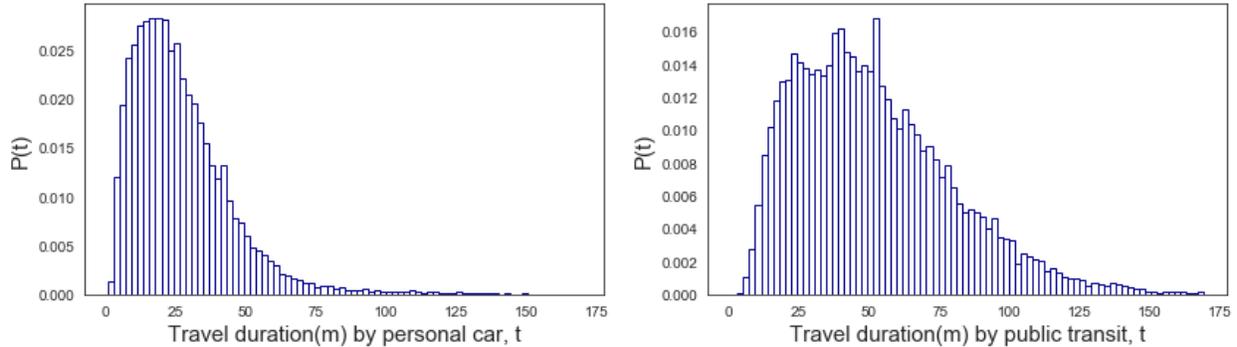

Figure. 4. (a) Home-to-work travel duration distribution using personal cars. (b) Home-to-work travel duration distribution for people in the sample for whom the engine could find a route using public transit.

**LIMITATION**

In this study, we constructed average daily home-work trips for 40,623 commuters in Houston and compared them to census data to validate the applicability of phone GPS data for estimating commute trips. While the results demonstrate a high correlation between GPS trips and CTTP trips, due to gaps in the city limits not all commuters who work in Houston were considered. Furthermore, we used two weeks of data to detect commuters and the sample should increase given longer periods of observation. Also, a longer observation period might result in sample workers distribution that is more representative of the population, and thus a better representative of trip distributions. Finally, although we used the most recent census data, there is still a gap of seven years with GPS data, and trips might have changed to some extent during this time.

**CONCLUSION**

In this study, we developed the algorithms to extract significant locations where the user has engaged in an activity within a specific timeframe. We investigated the effect of time thresholds used to extract these locations with an emphasis on avoiding oversimplification and computationally expensive tasks. This investigation is important because there are limited texts and research on ways of finding the optimal time threshold for extracting stay points. Given the higher temporal resolution of the phone GPS data, we found that the time threshold used for extracting stay points is highly dependent on the types of activities to be detected. Extracting stay points significantly reduced the volume of the raw phone GPS data and filtered out the noises. Then, we were able to detect home and work locations and to find commuters. Home-to-work average daily trips were constructed for commuters based on the observed working days for all individuals. For validation, home-to-work trip distributions inferred from GPS were compared with home-to-work flow tabulations of the 2006–2010 CTPP at tract-to-tract level. We observed a high correlation between the two datasets. The results confirm that phone GPS data can create a clearer picture of the spatial distribution of trips to address traffic demand issues.

ACKNOWLEDGMENT

This work was supported by National Science Foundation Grant HDBE-1761950; Northeastern Tier 1 Project on "Neighborhood Connectivity and Social Inequality"; and Global Resilience Institute Project on "Geosocial Network Resilience". To protect the confidentiality of any given individual's movement trajectory, all individuals' information from the phone GPS data was encrypted, and all data are reported in nonidentifiable form. All data used in this paper were reviewed and exempted by the Northeastern University Institutional Review Board (IRB). The data are proprietary and will not be shared.